\documentclass[twocolumn,aps,superscriptaddress,showpacs,nofootinbib]{revtex4}
\usepackage{hyperref}
\usepackage{footnote}
\usepackage{amssymb}
\usepackage{amsmath}
\usepackage{graphicx}
\usepackage[normalem]{ulem}
\usepackage{url}
\usepackage{csquotes}
\usepackage{color}

\begin{document}
\title{Isospin dependent hybrid model for studying isoscaling in  heavy ion collisions around the Fermi energy domain}

\author{S. Mallik}
\email{swagato@vecc.gov.in}
\affiliation{Physics Group, Variable Energy Cyclotron Centre, 1/AF Bidhan Nagar, Kolkata 700064, India}

\author{G. Chaudhuri}
\affiliation{Physics Group, Variable Energy Cyclotron Centre, 1/AF Bidhan Nagar, Kolkata 700064, India}
\affiliation{Homi Bhabha National Institute, Training School Complex, Anushakti Nagar, Mumbai 400085, India}


\begin{abstract}
Investigation of observables from nuclear multifragmentation reactions depending on isospin led to the development of a hybrid model. The mass and charge distribution as well as isotopic distribution was studied using this model for $^{112}$Sn+$^{112}$Sn reaction as well as $^{124}$Sn+$^{124}$Sn reactions at different energies. The agreement of the results obtained from the model with those from experimental data confirms the accuracy of the model. Isoscaling coefficients were extracted from these observables which can throw light on the symmetry energy coefficient. Another important facet of this model is that temperature of the studied reaction can be directly extracted using this model.

\end{abstract}

\maketitle
\section{Introduction}
The study of isospin dependent observables in nuclear multifragmentation reaction around the Fermi energy domain is a subject of contemporary interest \cite{Bao-an-li2,DasGupta_book}. Different statistical models like Statistical Multifragmentation Model (SMM) \cite{Bondorf1}, Canonical Thermodynamical Model \cite{DasGupta_book,Das1},  have been explored to investigate and verify the phenomenon of isoscaling \cite{Tsang5,Botvina1,Gargi1,Mallik5,Mallik8} as observed in experiments \cite{Tsang1,Souliotis,Shetty2,Fevre,Henzlova1,Shetty3}. The main motivation behind inclusion of isospin in the transport model based on Boltzmann-Uehling-Uhlenbeck (BUU) equation \cite{DasGupta_book,Dasgupta} was to study the isospin dependent observables in this framework. A hybrid model was developed  in our group few years back in order to study the central collision of Xe on Sn\cite{Mallik11} at beam energies around the Fermi energy domain and  the results were compared successfully with experimental data \cite{Hudan}. This work is an extension of the earlier one where isospin degree of freedom is incorporated in the transport model  based on the Boltzmann-Uehling-Uhlenbeck (BUU) approach. This distinction between neutron and proton  in the entrance channel enabled us to investigate isospin dependent observables in the exit channels. Good agreement with experimental data verified our approach (isospin dependent model) within a reasonable accuracy.\\
\indent
The models for explaining nuclear multifragmentation reactions can be broadly categorised under two heads. (i) Statistical Models \cite{Bondorf1,Das1,Gross2} and (ii) Dynamical Models \cite{Dasgupta,Ono1,Hartnack}. The statistical models which are based on phase space considerations has nice clusterization techniques included in them; the disadvantage of these models being that they assume some initial conditions like temperature, source size, freeze out volume etc. These conditions are either obtained from experimental observables or parameterized.  The dynamical models are based on  more microscopic calculations which deal with time evolution of the nucleons in phase space. These  models when coupled to the initial stage of the statistical models can fix these parameters from realistic considerations. Our present work is based on such hybrid model where the initial part of the nucleus nucleus collision is analysed using the BUU equation while the clusterization part is taken care of by the Canonical Thermodynamical model. The significance of this model is that one can extract the temperature of the studied reaction directly bypassing all ambiguities.\\
\indent
The excitation of the colliding system is calculated by using dynamical Boltzmann-Uehling-Uhlenbeck (BUU) approach \cite{Dasgupta,Mallik9} with appropriate consideration of pre-equilibrium emission. Then the disassembly of this excited system is analysed by Canonical Thermodynamical model (CTM) \cite{Das1}. The decay of excited fragments, which are produced in multifragmentation stage is calculated by the Weisskopf evaporation model \cite{Mallik1}. Charge, mass and isotopic distributions are the different observables which have been examined using this model for $^{112}$Sn+$^{112}$Sn reaction as well as $^{124}$Sn+$^{124}$Sn reaction  at 50MeV/nucleon \cite{Liu} subsequently compared with experimental data. We study central collisions around fermi energy domain which are extensively  used for producing neutron rich isotopes and for studying nuclear liquid gas phase transition. But in this particular work our main focus will be centered around the isospin dependent observables . More specifically we will examine the isotopic distributions of different elements from the two reactions and finally investigate if the well studied phenomenon of isoscaling emerges from our isospin dependent hybrid model. Isoscaling\cite{Botvina1,Gargi1,Mallik5,Mallik8,Tsang1,Souliotis,Shetty2,Fevre,Henzlova1,Shetty3,Ono2,Mallik102,Tsang_BUU1,Tsang_BUU2} is an important technique observed in certain reactions which depends crucially on the isospin of the system and thus can throw light on the symmetry energy\cite{Bao-an-li2}. This aspect has motivated both the the theoreticians and the experimentalists to study isoscaling and its relation with the symmetry energy coefficient. The present work is another effort in this direction. The fact that temperature can be estimated using this hybrid model facilitated the verification of similar temperature assumption made in the isoscaling equation\cite{Botvina1,Gargi1,Tsang1,Mallik8}. In the next section we will briefly describe our model and then present the results in the subsequent section.
\section{Basics of the model}
The theoretical calculation consists of three different stages: (i) Initial condition determination by isospin dependent Boltzmann-Uehling-Uhlenbeck model (BUU@VECC-McGill), (ii) fragmentation by canonical thermodynamical model and (iii) decay of excited fragments by evaporation model.\\
\indent
The BUU@VECC-McGill transport model calculation \cite{Mallik22,Zhang_code_comparison,Ono_code_comparison} for heavy ion collisions starts with two nuclei in their respective ground states approaching each other with specified velocities. For calculating ground state energies and densities an isospin dependent Thomas-Fermi model is developed separately which is briefly discussed in the Appendix section. The Thomas-Fermi phase space distribution is then sampled using  Monte-Carlo technique by choosing test particles (we use $N_{test}=100$ for each neutron or proton) with appropriate positions and momenta.\\
\indent
In the center of mass frame, the test particles of the projectile and the target nuclei 
are boosted towards each other. Simulations are done in a 200$\times 200\times 200 fm^3$ box. At t=0 fm/c the projectile and target nuclei are centered at (100 fm,100 fm,90 fm) and (100 fm,100 fm,110 fm). The test particles move in a mean-field $U(\rho_p(\vec{r}),\rho_n(\vec{r}))$ and occasionally suffer two-body collisions, with probability determined by the nucleon-nucleon scattering cross section. For each collision, Pauli blocking is checked and if the final states are allowed the momenta of the colliding particles are changed. The mean field potential used for this work is given by
\begin{eqnarray}
U(\rho_p(\vec{r}),\rho_n(\vec{r}))_{n/p}&=&a\rho(\vec{r})+b\rho^{\sigma}(\vec{r})\nonumber\\
&+&C_{sym}(\rho_n(\vec{r})-\rho_p(\vec{r}))\tau_z\nonumber\\
&+&c\nabla_r^2\rho(\vec{r})+\frac{1}{2}(1-\tau_z)U_c
\label{Mean_field_potential}
\end{eqnarray}
where $\rho(\vec{r})=\rho_p(\vec{r})+\rho_n(\vec{r})$; $\rho_n(\vec{r})$ and $\rho_p(\vec{r})$ are neutron and proton densities at $\vec{r}$ and $\tau_z$ is the zth component of the isospin degree
of freedom, which is $1$ or $-1$ for neutrons or protons respectively. First two terms in eq. \ref{Mean_field_potential} represent zero range Skyrme interaction, third term is due to isospin asymmetry, $U_c$ is the standard Coulomb interaction potential and the derivative term does not affect nuclear matter properties but in a finite system it produces quite realistic diffuse surfaces and liquid drop binding energies. This can be achieved for $A=$-2230.0 MeV $fm^3, B$=2577.85 MeV $fm^{7/6}, \sigma=$7/6, $\rho_0=0.16$ and $c$=-137.5 MeV$fm^{5}$ \cite{Lenk}. Co-efficient for isospin term is $C_{sym}$=200 MeV-$fm^3$. The mean-field propagation is done by using the lattice Hamiltonian Vlasov method which conserves energy and momentum very accurately \cite{Lenk,Mallik10}. Two body collisions are calculated as in Appendix B of ref. \cite{Dasgupta}, except that the pion channels are closed, as there will not be any pion production at 50 MeV/nucleon.\\
\indent
One can calculate the excitation energy from projectile beam energy by direct kinematics by assuming that the projectile and the target fuse together. In that case the excitation energy is too high as a measure of the excitation energy of the system which multifragments.  Pre-equilibrium particles which are not part of the multifragmenting system carry off a significant part of the energy. To get a better measure of excitation of the fragmenting system the pre-equilibrium particles can be identified after BUU simulation at the freeze-out stage and can be taken out. In different multifragmentation experiments, it is observed that after pre-equilibrium emission around $75\%$ to $80\%$ of the total mass creates the fragmenting system \cite{Xu,Frankland,Verde}. To make it consistent with the other exiting works on the $^{112}$Sn+$^{112}$Sn and $^{124}$Sn+$^{124}$Sn reactions at 50 MeV/nucleon, we choose the test particles which create $75\%$ of the total mass from the most central dense region.\\
\indent
At the end of the transport calculation in the freeze-out stage, the positions and momenta of the test particles forming central dense region are known; hence from these positions and momenta, one can calculate the potential and kinetic energies respectively. By adding kinetic and potential energy the excited state energy of the cluster can be obtained. It is observed that excited state energy become almost same for $t\geq 100$ fm/c \cite{Mallik11}. Therefore one can stop transport simulation at any time $t\geq 100$ fm/c and switch to statistical model. We have stopped the time evolution at $t=200$ fm/c However, to know the excitation one needs to calculate the ground state state energy also. This is done by applying the Thomas Fermi method for a spherical (ground state) nucleus having mass equal to the cluster mass. Then subtracting the ground state energy, the excitation is obtained. Knowing mass and excitation of the fragmentating system, the freeze-out temperature is calculated by using the canonical thermodynamic model CTM \cite{Das1,Mallik2,Mallik3}.\\
\begin{figure}[b]
\begin{center}
\includegraphics[width=0.8\columnwidth,keepaspectratio=true]{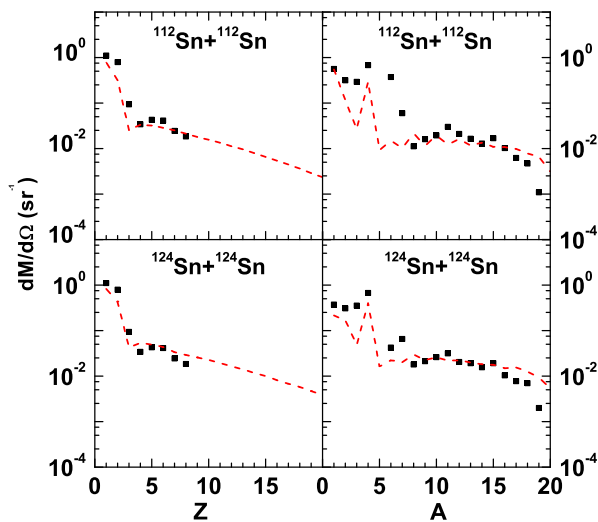}
\caption{Theoretical (red dashed lines) charge (left panels) and mass distribution (right panels) of $^{112}$Sn on $^{112}$Sn (upper panels) and $^{124}$Sn on $^{124}$Sn reaction (lower panels) reaction at 50 MeV/nucleon.The experimental data are shown by black squares.}
\end{center}
\end{figure}
\begin{figure}[t]
\begin{center}
\includegraphics[width=0.8\columnwidth,keepaspectratio=true]{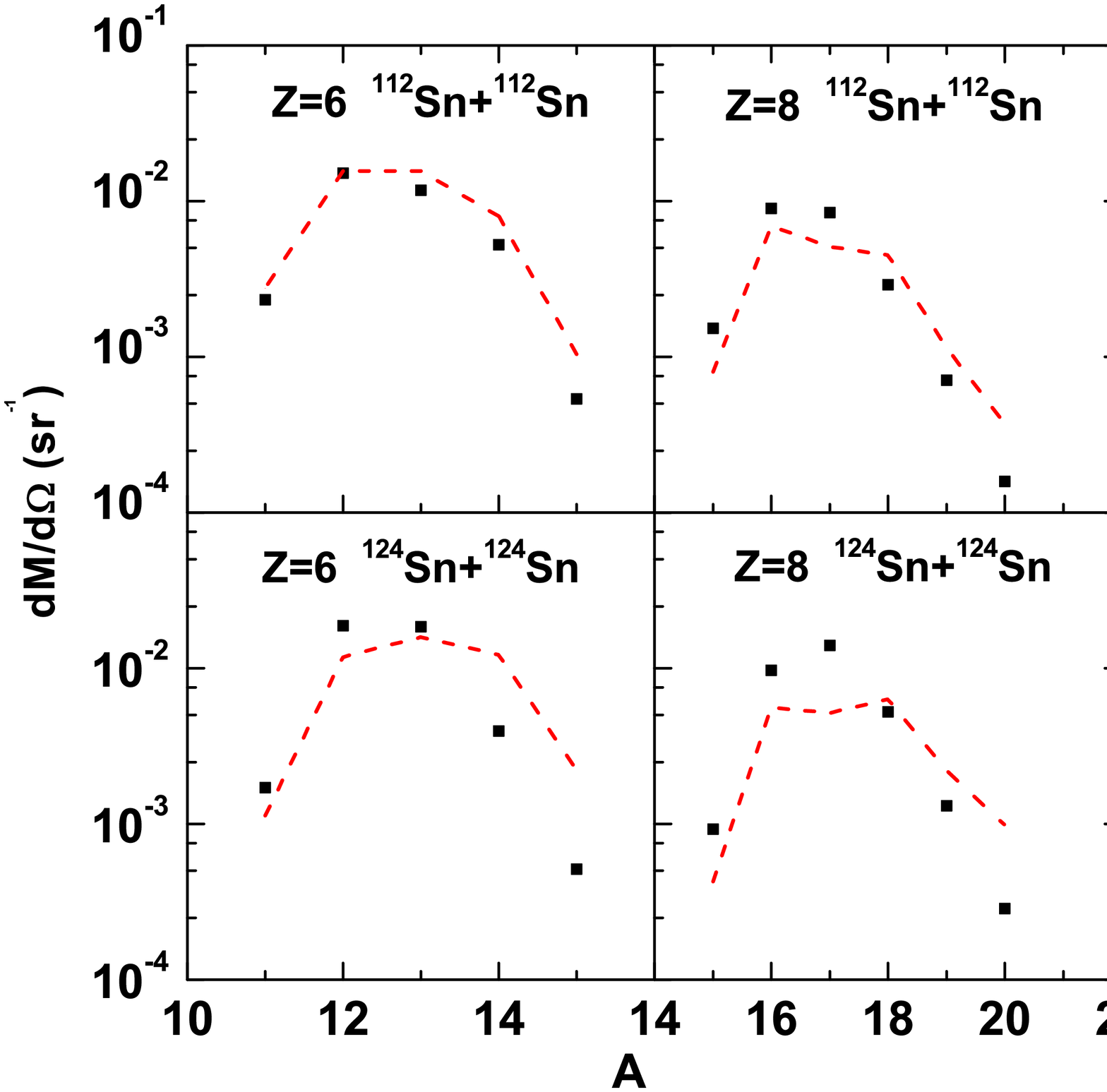}
\caption{Theoretical (red dashed lines) isotopic distributions (red dashed lines) at $Z$=6 (left panels) and 8 (right panels) for $^{112}$Sn on $^{112}$Sn (upper panels) and $^{124}$Sn on $^{124}$Sn reaction (lower panels) reaction reaction at 50 MeV/nucleon.The experimental data are shown by black squares.}
\end{center}
\end{figure}
\indent
Indeed CTM  can be used to calculate the average excitation per nucleon for a given temperature and mass number, and the relation is inversed to get the temperature from the output of the dynamical stage\cite{Mallik11,Mallik9,Mallik18}. In CTM, it is assumed that a system with $Z_0$ protons and $N_0$ neutrons at temperature $T$, has expanded to a higher than normal volume where the partitioning into different composites can be calculated according to the rules of equilibrium statistical mechanics. According to this model, the average number of composites with $N$ neutrons and $Z$ protons can be calculated from,
\begin{equation}
\langle n_{N,Z}\rangle = \omega_{N,Z}\frac{Q_{N_{0}-N,Z_{0}-Z}}{Q_{N_{0},Z_{0}}}
\end{equation}
where, $\omega_{N,Z}$ is the partition function of one composite with $N$ neutrons and $Z$ protons and $Q_{N_0,Z_0}$ is the total partition function which can be calculated from the recursion relation,
\begin{equation}
Q_{N_0,Z_0}=\frac{1}{N_0}\sum_{N,Z}N\omega_{N,Z}Q_{N_0-N,Z_0-Z}
\end{equation}\\
The description of $\omega_{N,Z}$ and details of CTM can be found in Ref. \cite{Das1}.\\
\indent
The excited fragments produced after multifragmentation decay to their stable ground states. They can $\gamma$-decay to shed energy but may also decay by light particle emission to lower mass nuclei.  We include emissions of $n,p,d,t,^3$He and $^4$He. Particle-decay widths are obtained using Weisskopf's evaporation theory. Fission is also included as a de-excitation channel though for the nuclei of $A{\textless}100$ its role will be quite insignificant. The details of the evaporation stage are described in Ref. \cite{Mallik1}.\\
\section{Results}
We have done calculations for the $^{112}$Sn+$^{112}$Sn and $^{124}$Sn+$^{124}$Sn reaction for projectile beam energy 50 MeV/nucleon. In Fig. 1 we have plotted the charge (left) and mass (right) distributions of the fragments. In the given excitation energy regime, the distribution decreases with charge or mass number. Results have been compared with experimental data and good agreement has been obtained as is visible from the figure. This establishes the success of our model where the initial part has been described by the isospin dependent BUU equation followed by the canonical thermodynamical model(CTM) for the deexcitation part. This motivated us to further probe into the details and thereby study the isotopic distribution of some elements. In Fig. 2 have plotted the isotopic distribution of carbon and oxygen  from $^{112}$Sn+$^{112}$Sn (upper panels) and $^{124}$Sn+$^{124}$Sn (lower panels) reactions. The production cross-section varies widely over orders of magnitude and our model could successfully predict this large change. The behaviour as seen for $^{124}$Sn+$^{124}$Sn reaction is pretty similar as that of $^{112}$Sn+$^{112}$Sn , the difference being that the cross section of neutron rich isotopes are more for the neutron rich reaction for obvious reasons. Here too the model calculation could do good justice to the experimental data.\\
\indent
In Fig.3 we  have displayed the isoscaling which is the ratio of the yields of the same fragment from the neutron rich to that of the neutron less isotope reaction. In the left panel, the odd Z ones are plotted while the even ones are plotted in the right panel just for the sake of clarity. The plots are really nice with the lines approximately parallel to each other as it should be if the law of isoscaling is obeyed. The slope of these parallel lines is somewhat related to the symmetry energy coefficients $\alpha$ and $\beta$  as given by the following isoscaling equation.
\begin{eqnarray}
R_{21}(N,Z)&=& \langle {n_2}_{N,Z}\rangle/\langle {n_{1}}_{N,Z}\rangle\nonumber\\
&=& C\exp(\frac{\mu_{n_2}-\mu_{n_1}}{T}N+\frac{\mu_{z_2}-\mu_{z_1}}{T}Z)\nonumber\\
&=& C\exp(\alpha N+\beta Z)
\end{eqnarray}
\begin{figure}[t]
\begin{center}
\includegraphics[width=0.95\columnwidth,keepaspectratio=true]{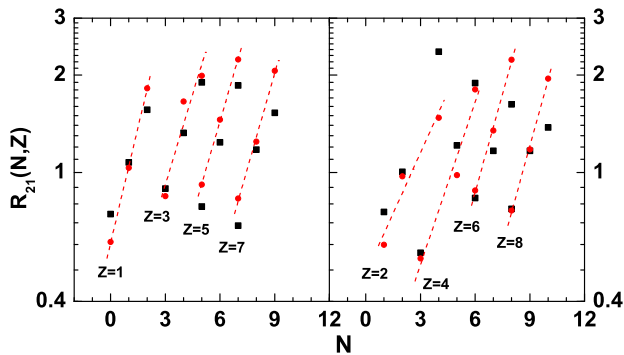}
\caption{Isotopic ratios($R_{21}$) of multiplicities of fragments $(N,Z)$ where reaction 1 and 2 are $^{112}$Sn on $^{112}$Sn and $^{124}$Sn on $^{124}$Sn respectively. For both reaction the projectile beam energy is 50 MeV/nucleon. The left panel shows the ratios as function of neutron number $N$ for fixed $Z$ values, while the right panel displays the ratios as function of proton number $Z$ for fixed neutron numbers ($N$). The red dashed lines are drawn through the best fits of the theoretically calculated ratios (red circles). The experimental data are shown by black squares.}
\end{center}
\end{figure}
\begin{table}[b]
\begin{center}
\begin{tabular}{|c|c|c|}
\hline
Isoscaling& Theoretical& Experimental\\
Parameter& &\\
\hline
$\alpha$ & 0.41 & 0.36\\
\hline
$\beta$ & -0.49 & -0.42\\
\hline
\end{tabular}
\end{center}
\caption{Best fit values of the isoscaling parameters $\alpha$ and $\beta$ for the two reactions $^{112}$Sn on $^{112}$Sn and $^{124}$Sn on $^{124}$Sn. The values obtained from the slope of the primary and secondary fragments as well as the experimental values are tabulated. The theoretical values of $\alpha$ and $\beta$ are extracted for fragments having $Z\leqslant 8$ and $N\leqslant 8$ respectively.}
\label{table1}
\end{table}

where $R_{21}$ is the ratio of the yields from two reactions and $C$ is some arbitrary constant, $\mu_{n}$'s and $\mu_{z}$'s are neutron and proton chemical potential of the two fragmenting sources at freeze-out condition. One vital assumption made while deriving the above equation  is that freeze-out temperature ($T$) of both the reactions are same. The concept of temperature is quite familiar in heavy ion physics and it is usually calculated \cite{Dasgupta_Phase_transition,Pochodzalla,Agrawal,Trautmann_temperature,Mallik7} from double isotope ratio method \cite{Albergo} or kinetic energy spectra of emitted particles. But in both cases, sequential decay from higher energy states \cite{Nayak2}, Fermi motion \cite{Bauer2}, pre-equilibrium emission etc complicate the scenario of temperature measurement and the response of different thermometers is sometimes contradictory \cite{Xi_temperature,Francesca_temperature}. The advantage of using this hybrid model calculation is that one can estimate the temperature of the intermediate energy heavy ion reactions  directly from here which bypasses all such problems. It is directly obtained from this isospin dependent hybrid model calculation that, $T$ for the $^{112}$Sn+$^{112}$Sn reaction is 5.04 MeV while that for the $^{124}$Sn+$^{124}$Sn reaction is 5.08 MeV. The values are extremely close and this confirms strongly the assumption made for applying isoscaling equation. The fact that temperature can be directly calculated from this model enabled the testing as well as verification of this assumption.\\
\indent
The linear fits to the data points as obtained from our model as well as experimental data can be used to extract the values of $\alpha$ and $\beta$ which have been tabulated (Table 1). The closeness of the values extracted from our model with those from the experiment establishes the validity of our model.\\
\section{Discussions}
The study of isospin dependent observables,  more specifically isoscaling is done by hybrid model. The dynamical approach takes care of the initial stages of the reaction and the fragmentation of the excited system is described by the statistical model. This hybrid model is much economical but at the same time based on appropriate physical considerations at different stages of the reaction. The introduction of isospin in the model led us to study the observables dependent on isospin in order to check the accuracy of the model. Nice fits to the experimental data of isospin distribution from the two reactions confirms the validity of the model. Isoscaling is also nicely displayed by the ratios of yields from the two reactions and the straight line fits to the same have been used to extract the isoscaling coefficients. One significance of using this hybrid model is direct estimation of temperature of the reactions being studied. This feature enabled us to confirm the important assumption of isoscaling equation, that is, temperature of both the reactions are very close.  The satisfactory performance of the model motivates us to use it in order to probe experimental data of other isospin dependent observables in future. This model can also be extended in future in order to study projectile fragmentation reactions in the higher energy domain.\\
\section{Appendix: Isospin dependent Thomas-Fermi Model}
Consider a nucleus of $Z_0$ protons and $N_0$ neutrons. The total energy (non-relativistic) of the system for mean field given in eq. \ref{Mean_field_potential} can be expressed as
\begin{eqnarray}
E&=&\frac{3h^2}{10m}\bigg{[}\frac{3}{8\pi}\bigg{]}^{2/3}\Bigg{\{}\int \rho_p(r)^{5/3}d^3r+\int \rho_n(r)^{5/3}d^3r\Bigg{\}}\nonumber\\
&+&\frac{a}{2}\int \rho^2(r)d^3r+\frac{b}{\sigma+1}\int \rho^{\sigma+1}({r})d^3r\nonumber\\
&+&\frac{c}{2}\int \rho(\vec{r})\nabla_r^2\rho(r)d^3r+\frac{C_{sym}}{2}\int (\rho_n(r)-\rho_p(r))^2d^3r\nonumber\\
&+&\frac{1}{4\pi\epsilon_0}\int\int \frac{\rho_p(\vec{r})\rho_p(\vec{r'})}{|\vec{r}-\vec{r'}|} d^3rd^3r'
\label{Energy}
\end{eqnarray}
And the particle number conservation gives,
\begin{eqnarray}
\int \rho_p(r)d^3r=Z_0\nonumber\\
\int \rho_n(r)d^3r=N_0
\label{Total_Proton_&_Neutron}
\end{eqnarray}
By applying the variational method of energy minimization under the constraint of total proton and neutron conservation and assuming spherical symmetry, one can get the two Thomas-Fermi equations
\begin{eqnarray}
\frac{h^2}{2m}\bigg{[}\frac{3}{8\pi}\bigg{]}^{\frac{2}{3}}\rho_p^{\frac{2}{3}}(r)+\Big{\{}a\rho(r)+b\rho^{\sigma}(r)+c\nabla_r^2\rho(r)\Big{\}}\nonumber\\
-C_{sym}\Big{\{}\rho_p(r)-\rho_n(r)\Big{\}}+\frac{1}{4\pi\epsilon_0}\int \frac{\rho_p(\vec{r'})}{|\vec{r}-\vec{r'}|} d^3r'-\lambda_p&=&0\nonumber\\
\label{Proton_Thomas-Fermi}
\end{eqnarray}
\begin{eqnarray}
\frac{h^2}{2m}\bigg{[}\frac{3}{8\pi}\bigg{]}^{\frac{2}{3}}\rho_n^{\frac{2}{3}}(r)+\Big{\{}a\rho(r)+b\rho^{\sigma}(r)+c\nabla_r^2\rho(r)\Big{\}}\nonumber\\
+C_{sym}\Big{\{}\rho_p(r)-\rho_n(r)\Big{\}}-\lambda_n&=&0\nonumber\\
\label{Neutron_Thomas-Fermi}
\end{eqnarray}
where $\lambda_p$ and $\lambda_n$ are Lagrange undetermined multiplier for proton and neutron conservation respectively. To solve these two coupled equations simultaneously, one can put $y_p(r)=r\rho_p(r)$ and $y_n(r)=r\rho_n(r)$, so $y_p(r)$ and $y_n(r)$ vanishes both at $r=0$ and $r=\infty$ i.e. above two coupled equations become boundary value problem.\\
\indent
Numerically one have to start from a guess proton and neutron density profile (for example, we have started with Myers density profile\cite{Myers}) and guess value of $\lambda_p$ and $\lambda_n$ and by applying multidimensional Newton's method (at different $r$ values) for coupled  equations \ref{Proton_Thomas-Fermi} and \ref{Neutron_Thomas-Fermi} the ground state neutron and proton density profile can be obtained. Then by using Monte-Carlo technique the initial position and momenta of the proton and neutron test particles as well as ground state energy can be obtained from the calculated ground state density profile.\\

\end{document}